%% file: main.tex
\documentclass[sigconf, nonacm]{acmart}
\renewcommand\footnotetextcopyrightpermission[1]{}
\renewcommand\keywords[1]{}

\AtBeginDocument{%
  }

\setcopyright{acmlicensed}
\copyrightyear{2018}
\acmYear{2018}
\acmDOI{XXXXXXX.XXXXXXX}
\acmISBN{978-1-4503-XXXX-X/2018/06}

\usepackage{amsmath}
\usepackage{amsfonts}
\usepackage{graphicx}
\usepackage{algorithmic}
\usepackage{multirow}
\usepackage{booktabs}
\usepackage{makecell}
\usepackage{textcomp}
\usepackage{colortbl} 
\usepackage{arydshln}
\usepackage{geometry}
\usepackage{array}
\usepackage{relsize}
\usepackage[dvipsnames]{xcolor}
\usepackage{listings} 
\usepackage{arydshln}

\usepackage[status=final,nomargin,inline]{fixme}
\usepackage{placeins}
\fxusetheme{colorsig}
\FXRegisterAuthor{dg}{adg}{\textcolor{green}{by DG}}
\FXRegisterAuthor{td}{atd}{\textcolor{green}{ToDo}}
\FXRegisterAuthor{hw}{ahw}{\textcolor{red}{for HW Team}}
\FXRegisterAuthor{ma}{mag}{\textcolor{blue}{MA}}
\FXRegisterAuthor{bh}{bhg}{\textcolor{purple}{by BH}}
\FXRegisterAuthor{ep}{epg}{\textcolor{cyan}{EP}}
\FXRegisterAuthor{gk}{gkg}{\textcolor{magenta}{by GK}}
\FXRegisterAuthor{rmg}{rmg}{\textcolor{BurntOrange}{Razine}}

\usepackage{url} 
\usepackage{xurl}         

\usepackage{xspace}
\usepackage{textcase}

\usepackage{hyperref}

\newcommand{\leaderboardurl}{\texttt{https://huggingface.co/spaces/HPAI-BSC/TuRTLe-Leaderboard}}

\newcommand{\dataseturl}{\texttt{https://huggingface.co/datasets/HPAI-BSC/NotSoTiny-25-12}}
\usepackage{soul}

\makeatletter
\def\@ACM@checkaffil{} 
\makeatother

\begin{document}

\input{sections/0_Header}
\input{sections/1_Introduction}
\input{sections/2_NotSoTiny_Benchmark_Construction}
\input{sections/3_LivingBench}
\input{sections/4_Evaluations}
\input{sections/5_Conclusions}

\input{sections/acknowledgment}


\bibliographystyle{ACM-Reference-Format}
\bibliography{references}

\appendix{}
\label{sec:Appendix}
\input{sections/appendix.tex}

\end{document}

%% file: sections/0_Header.tex
\title{NotSoTiny: A Large, Living Benchmark for RTL Code Generation}

\author{Razine Moundir Ghorab}
\email{moundir.ghorab@bsc.es}
\orcid{https://orcid.org/0009-0005-2541-6867}
\affiliation{
   \institution{\textit{Barcelona Supercomputing Center}}
 }
\author{Emanuele Parisi}
\email{emanuele.parisi@bsc.es}
\orcid{https://orcid.org/0000-0001-6607-7367}
\affiliation{
   \institution{\textit{Barcelona Supercomputing Center}}
 }
\author{Cristian Gutierrez-Gomez}
\email{cristian.gutierrez@bsc.es}
\orcid{https://orcid.org/0009-0005-1441-8568}
\affiliation{
   \institution{\textit{Barcelona Supercomputing Center}}
 }
\author{Miquel Alberti-Binimelis}
\email{miquel.alberti@bsc.es}
\orcid{https://orcid.org/0009-0005-1666-8421}
\affiliation{
   \institution{\textit{Barcelona Supercomputing Center}}
 }
\author{Miquel Moreto}
\email{miquel.moreto@bsc.es}
\orcid{https://orcid.org/0000-0002-9848-8758}
\affiliation{
   \institution{\textit{Barcelona Supercomputing Center}}
   \institution{\textit{Universitat Politecnica de Catalunya}}
 }
\author{Dario Garcia-Gasulla}
\email{dario.garcia@bsc.es}
\orcid{https://orcid.org/0000-0001-6732-5641}
\affiliation{
   \institution{\textit{Barcelona Supercomputing Center}}
 }
\author{Gokcen Kestor}
\email{gokcen.kestor@bsc.es}
\orcid{https://orcid.org/0000-0002-9105-5634}
\affiliation{
   \institution{\textit{Barcelona Supercomputing Center}}
 }


\begin{abstract}
LLMs have shown early promise in generating RTL code, yet evaluating their capabilities in realistic setups remains a challenge. So far, RTL benchmarks have been limited in scale, skewed toward trivial designs, offering minimal verification rigor, and remaining vulnerable to data contamination. To overcome these limitations and to push the field forward, this paper introduces NotSoTiny, a benchmark that assesses LLM on the generation of structurally rich and context-aware RTL. Built from hundreds of actual hardware designs produced by the Tiny Tapeout community, our automated pipeline removes duplicates, verifies correctness and periodically incorporates new designs to mitigate contamination, matching Tiny Tapeout release schedule. Evaluation results show that NotSoTiny tasks are more challenging than prior benchmarks, emphasizing its effectiveness in overcoming current limitations of LLMs applied to hardware design, and in guiding the improvement of such promising technology.
\end{abstract}



\keywords{LLM-assisted circuit design, Electronic Design Automation}


\maketitle

%% file: sections/1_Introduction.tex
\section{Introduction}
\label{sec:Introduction}

The use of Large Language Models (LLMs) for RTL code generation has recently attracted considerable interest in the hardware design automation community. Thanks to their strong code-generation capabilities, LLMs can translate high-level specifications, including natural-language descriptions, diagrams, and tables, into Verilog, potentially accelerating and partially automating digital hardware development. However, despite this growing promise, progress is limited by existing benchmarks, which fail to accurately reflect the complexity, rigor, and practical challenges of real-world hardware design workflows~\cite{pan2025survey,he2025large}.

Current RTL generation benchmarks suffer from three main limitations: 1) insufficient scale and complexity, 2) shallow verification protocols, and 3) lack of contamination control. 
First, existing RTL generation benchmarks consist mainly of flat, narrowly scoped modules that lack the hierarchical composition and design diversity found in real-world hardware. Practical digital systems, in contrast, feature deep module hierarchies and complex control and datapath interactions. This structural gap restricts the ability of current benchmarks to guide the development of LLMs with strong generalization and reasoning capabilities. Additional evidence of limited benchmark complexity is the consistently high pass rates achieved by current models on existing benchmarks~\cite{thakur2024verigen,liu2023verilogeval,lu2024rtllm}, which provides a weak signal of model progress, and fail to reflect the challenges posed by realistic RTL code.

As a second main limitation, existing evaluation frameworks often apply minimal verification rigor; in most cases, LLM-generated designs are only checked for syntax validity and exercised using predefined simulation testbenches. While testbenches offer a straightforward means of assessing a module’s behavior, they may exhibit insufficient coverage, overlook corner cases, or implement too few checks to be considered trustworthy. As a result, such evaluations often miss subtle behavioral mismatches, particularly in designs featuring complex control logic, or deep datapaths.


The third issue of current benchmarks, contamination of evaluation data leaking into the training set, is often overlooked. Many benchmark designs are sourced from publicly available, well-known textbooks, open source repositories and educational sites. These resources are among the most commonly used ones to train modern LLMs, enabling these models to simply reproduce partially memorized code rather than generating novel solutions. Furthermore, most prior benchmarks for RTL code generation include no contamination mitigation strategy, making it difficult to distinguish genuine generalization from recall. Recent studies have demonstrated the severity of this issue in popular code benchmarks (e.g., HumanEval~\cite{chen2021evaluatinglargelanguagemodels}), which have been found to be present in the training corpora of production models like ChatGPT, significantly inflating performance metrics~\cite{dong2024datacontamination}.

To address these limitations, this work presents NotSoTiny, a new benchmark for evaluating LLMs on structurally complex, context-aware RTL code-generation tasks, explicitly designed to be resilient to contamination. NotSoTiny is initially built from hundreds of hardware designs contributed by the Tiny Tapeout community, an open, educational multi-project wafer program where researchers and engineers submit digital, mixed signal, and analog circuits for fabrication~\cite{tinytapeout}. These designs span a wide range of domains, styles, and abstraction levels, including hierarchical systems, datapath-heavy logic, finite-state machines, and custom protocols. After curating the TinyTapout projects, we automatically generate over 1,000 \textit{contextual-module completion} tasks, an order of magnitude larger than the next-largest RTL benchmark, where each task is derived from a real, taped out hardware design. On each task, one module is withheld, and LLMs are prompted to regenerate the missing component using the remaining design as context. Unlike prior benchmarks that provide explicit I/O specifications or standalone modules, NotSoTiny requires the model to infer functionality and interface behavior solely from the surrounding implementation. This mirrors real-world development scenarios where new components must integrate seamlessly into existing systems. 

To construct contextual-module completion tasks, we developed a fully automated pipeline that crawls Tiny Tapeout shuttle repositories. Each project is parsed to extract Verilog modules and testbenches, producing multiple completion tasks per design by “masking out” individual modules and treating them as the generation targets. To avoid duplication, we apply hash-based and structural similarity checks to identify and remove redundant designs, which are often re-submitted across shuttles with minor changes. This ensures that each circuit concept appears only once in the benchmark. Finally, each task is paired with its full design context (all remaining modules and top-level logic) and the corresponding ground-truth implementation of the withheld module. Outputs generated by LLMs are then evaluated using both simulation-based and formal verification techniques.


Experiments show that NotSoTiny is significantly more challenging than existing RTL benchmarks: state-of-the-art LLMs achieve only ~20\% functional correctness under formal equivalence, despite high syntax. Min-K analysis~\cite{min-k} further shows low contamination risk, establishing NotSoTiny as a rigorous and reliable benchmark. This leads to the following contributions:
 \begin{itemize}
     \item \textbf{A large and structurally rich RTL benchmark.} We introduce NotSoTiny, a dataset of 1,114 deduplicated module-completion tasks that is significantly larger and more complex than existing RTL benchmarks. The dataset is publicly available on HuggingFace\footnote{\dataseturl}.
     \item \textbf{A rigorous and scalable evaluation methodology.} We show that syntax checks and testbenches substantially overestimate model performance, and provide a formal verification pipeline based on equivalence checking that measures functional correctness at scale.
     \item \textbf{A contamination-resilient, periodically updated benchmark pipeline.} We develop an end-to-end construction pipeline, including crawling, merging, task generation, and near-duplicate filtering, that allows NotSoTiny to be refreshed with each new Tiny Tapeout shuttle, enabling long-term contamination resilience.
 \end{itemize}

%% file: sections/2_NotSoTiny_Benchmark_Construction.tex
\section{NotSoTiny Benchmark Construction}
\label{sec:NotSoTiny_Benchmark_Construction}

\begin{figure*}[t]
    \centering
    \includegraphics[width=1.0\textwidth]{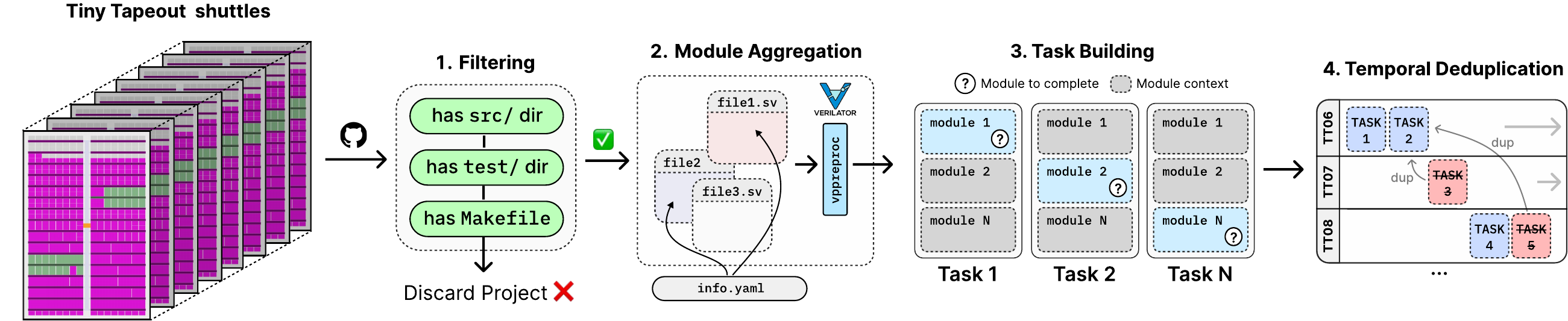}
    \caption{Overview of the NotSoTiny benchmark construction pipeline. Designs from multiple Tiny Tapeout shuttles are filtered, merged, converted into contextual-module completion tasks, and deduplicated to produce a clean, diverse, and contamination-resilient benchmark.}
     \label{fig:build-overview}
\end{figure*}

NotSoTiny is a novel benchmark for systematically evaluating LLM capabilities in RTL code generation, focusing on module completion. The benchmark is built from data derived from the Tiny Tapeout project, a collaborative initiative that allows hardware designers to submit open-source digital circuit designs for fabrication on periodic and independent shuttles~\cite{tinytapeout}. Each shuttle aggregates multiple individual designs, collectively forming a rich and diverse corpus of open-source digital hardware implementations. By means of these shuttles, Tiny Tapeout provides a periodic set of fresh designs that help mitigate potential contamination issues. To construct a reliable and representative benchmark based on Tiny Tapeout, we design a multi-stage data processing pipeline, illustrated in Figure~\ref{fig:build-overview}, comprising four main components: (1) Filtering, (2) Module Aggregation, (3) Task Building, and (4) Deduplication.

\subsection{Filtering and Module Aggregation}

The first stage of the benchmark construction pipeline involves curating and preprocessing raw RTL designs from Tiny Tapeout. For the initial release of NotSoTiny, a total of 1,572 projects from publicly accessible GitHub repositories were collected, associated with the fabrication shuttles \textit{TT06}, \textit{TT07}, \textit{TT08}, \textit{TT09}, \textit{TT10 IHP-02}, \textit{TT10 IHP-25a}, and \textit{TTsky25a}. Each project typically includes source code, tests and metadata describing its structure and build process. To ensure quality and consistency, we apply a \textit{Filtering} step that discards incomplete or malformed projects. Specifically, we retain only those projects that contain a \texttt{src/} directory with RTL source files, a \texttt{test/} directory with at least one testbench, a \texttt{Makefile} to build and run the design, and a valid \texttt{info.yaml} file that explicitly specifies all required source dependencies. Projects that do not satisfy these structural requirements, or for which referenced source files cannot be located, are excluded from further processing. After this filtering step, a total of 1,034 valid projects remain.

For each accepted project, the \textit{Module Aggregation} step unifies its RTL code into a single, stand-alone Verilog file, enabling uniform parsing and task generation. This step uses \texttt{vppreproc}~\cite{Snyder_Verilator}, a Verilog preprocessor from the Verilog-Perl suite, which resolves \textit{include} directives and macro definitions across multiple source files. The result is a fully expanded, self-contained Verilog file that preserves internal module hierarchies and dependencies for the total number of 1,021 projects. These files serve as the foundation for our contextual module completion tasks.

\subsection{Task Building}
The next stage decomposes each merged project file into a set of module completion tasks. If a merged file contains \textit{n} modules, then \textit{n} distinct tasks are generated from it. For each task, the body of a single module is removed and designated as the \textbf{target} for completion, while the remaining modules in the file are retained as \textbf{context}. This procedure yields a prompt that presents the model with a realistic design context and asks it to produce the missing module’s code. These task mimics the scenario where part of a hardware design is missing and must be implemented to integrate with the existing components.
In total, this stage yields 5,309 contextual-module completion tasks before deduplication.

\subsection{Temporal Deduplication}
The fourth stage eliminates redundant designs across Tiny Tapeout shuttles. Given the collaborative and iterative nature of the Tiny Tapeout project, contributors frequently resubmit existing designs, sometimes with minor modifications, or reuse modules from earlier shuttles. For example, a simple 8-bit counter module might be submitted by multiple people across different shuttle runs. Without proper filtering, such repetition would lead to artificially inflated evaluation metrics or unfairly advantage models that have seen one shuttle’s data during training. 
To address this, we adopt a near-duplicate detection pipeline inspired by the BigCode deduplication approach~\cite{bigcode-dedup} using Datasketch~\cite{eric_zhu_2017_290602}. Our deduplication framework combines MinHash fingerprinting with Locality-Sensitive Hashing (LSH)~\cite{Leskovec_Rajaraman_Ullman_2014} to efficiently identify similar RTL designs. This process consists of three sub-steps.

\paragraph{\textbf{Signature Generation via MinHash:}} Each merged RTL design, represented as a unified module-level artifact, is tokenized into overlapping world-level $k$-grams (shingles) that capture local code structure and syntax. 
Using a word shingle size of 5 (approximately 20 characters) to represent each design’s content, we compute MinHash signatures to obtain compact similarity-preserving representations of each design. The MinHash yields an approximate Jaccard similarity between any two designs in linear time.

\paragraph{\textbf{Candidate Identification via LSH:}} LSH is used to find candidate duplicates without exhaustive pairwise comparison. We partition each MinHash signature into \textit{b} bands and hash each band to an identifier. If two designs share at least one identical band hash, they are flagged as candidate near-duplicates. 
This banding technique reduces comparisons by only focusing on likely similar pairs. We configure the LSH such that it is expected to detect pairs with Jaccard similarity above a threshold T = 0.70, a value chosen to balance precision and recall for code duplication (similar to BigCode deduplication settings ~\cite{bigcode-dedup}).
Each candidate pair identified by LSH is then measured for actual Jaccard similarity to confirm the duplicate. If the similarity exceeds the threshold, we treat those designs as near-duplicates.

\paragraph{\textbf{Temporal Deduplication Policy}} After identifying near-duplicates, we apply a time-aware filter to decide which version of a duplicated design to keep. Specifically, within each set of duplicate tasks, we retain only the version from the oldest Tiny Tapeout shuttle and remove all later re-submissions. This policy ensures that the benchmark preserves the original instance of a design while eliminating subsequent copies. This incremental deduplication process lets the benchmark evolve with new shuttles without introducing duplicate content that a model might have already seen in prior versions.
After performing the deduplication stage, we end up with 3062 unique module-completion tasks from all Tiny Tapeout shuttles, each derived from a distinct hardware design and free of redundancy. Each task encapsulates a realistic RTL design scenario with contextual modules and a missing module to be completed, providing a challenging and representative testbed for evaluating LLM performance in automatic RTL code generation.


\subsection{Ensuring Compatibility via Self-Verification}

To finalize the dataset and complete the pipeline execution, we do a final filtering on the 3,062 unique tasks identified in the previous temporal deduplication step. Aligning with our objective to create a fully automatic benchmark with negligible manual intervention, we employed formal equivalence checking to verify the designs.

Our implementation utilizes a Yosys ~\cite{wolf2013yosys} script, the methodology of which is detailed in \S\ref{subsec:formalverif}. To ensure compatibility, the task's golden solution is verified against itself using this general verification script, rather than against an LLM-generated solution. This step filters for tasks that function 'out of the box' with the general script, eliminating the need for manual fine-tuning. Consequently, this validation step reduced the initial 3,062 tasks to a final benchmark dataset of 1,114 validated tasks.

\subsection{Benchmark Scale and Complexity}

The automated construction pipeline described above, including filtering, module aggregation, task building, and temporal deduplication, results in a benchmark that is not only clean and contamination-resilient but also substantially larger and more structurally diverse than existing RTL datasets. This is important because one of the central limitations of prior RTL benchmarks is their small scale and lack of architectural depth, which prevents them from capturing the complexity of real-world hardware design. Table \ref{tab:intro_stats} compares NotSoTiny against several state-of-the-art RTL benchmarks in terms of average sample complexity. We report key indicators, including lines of code (LOCs), and a structural complexity score computed by summing the number of Verilog keywords commonly used to describe module functionality (e.g., always, assign, generate, wire, reg) per design. NotSoTiny contains 1,114 deduplicated tasks, an order of magnitude more than previous datasets, and shows significantly higher structural complexity, averaging 118 LOCs per task, and a complexity score of 21.90, highlighting how the average sample in NotSoTiny is approximately 3 times more complex than the other considered benchmarks. These results suggest that NotSoTiny provides a considerably more challenging dataset for evaluating the performance of LLM on RTL code generation tasks. 

RealBench~\cite{jin2025realbench} contains samples with high LOC and complexity scores, as it comprises large IPs from open-source repositories. These designs boost the complexity metric through their size, but do not require models to reason about missing components within a broader system. In contrast, NotSoTiny derives its complexity from multi-module, context-dependent reasoning, where the model must reconstruct a missing module that integrates correctly with the existing hierarchy, evaluating the LLM's capacity to infer the logic to be implemented from the context, rather than from a specification document. 
This form of structural and semantic complexity aligns more closely with real engineering workflows.
Notice the methodology used to construct NotSoTiny is fully generalizable: it can be applied to any open-source hardware project, including those included in RealBench~\cite{jin2025realbench}, to generate arbitrarily complex, context-aware code-completion tasks, thus providing a scalable path for building future benchmarks that stress deeper reasoning capabilities in LLMs. Finally, the large number of tasks in NotSoTiny enables a more statistically robust evaluation than in static datasets with limited sample sizes.

\input{tables/intro_stats}

%% file: tables/intro_stats.tex
\begin{table}[t]
    \centering
    \caption{Comparison of NotSoTiny with existing RTL benchmarks in terms of scale and structural complexity. NotSoTiny provides significantly more samples and higher structural richness than prior datasets, while also being the only benchmark designed to be continuously updated (“living”).
    }
    \label{tab:intro_stats}

    \begin{tabular}{@{}lrrrr@{}}
        \toprule

        \multicolumn{1}{c}{\multirow{2}{*}{\vspace{-.2cm}Benchmark}} &
        \multicolumn{1}{c}{\multirow{2}{*}{\vspace{-.2cm}Samples}}   &
        \multicolumn{2}{c}{Sample statistics}                        &                       
        \multicolumn{1}{c}{\multirow{2}{*}{\vspace{-.2cm}Living}}    \\
        \cmidrule(lr){3-4}
                                         &
                                         &
        \multicolumn{1}{c}{LOCs}         &
        \multicolumn{1}{c}{Complexity}   &
                                         \\

        



        \midrule
        
        VeriGen~\cite{thakur2024verigen}            &    17 &  17.05 &  2.12 &  No \\
        VerilogEval~\cite{pinckney2025revisiting}   &   158 &  16.58 &  2.22 &  No \\
        RTLLM~\cite{lu2024rtllm}                    &    53 &  37.73 &  6.01 &  No \\
        CVDP cid02~\cite{pinckney2025comprehensive} &    94 &  71.09 &  6.52 &  No \\
        RealBench~\cite{jin2025realbench}           &    67 & 192.03 & 50.90 &  No \\

        \midrule

        NotSoTiny                                   & 1,114 & 118.07 & 21.90 & Yes \\

        \bottomrule
    \end{tabular}

\end{table}

%% file: sections/3_LivingBench.tex
\section{Living Benchmark}
\label{sec:livingbench}



A critical yet often overlooked limitation of existing RTL benchmarks is their vulnerability to training-data contamination. As LLMs increasingly ingest publicly available datasets, including textbooks, open-source repositories, and educational materials, there is a growing concern that benchmark tasks may overlap with the model’s pretraining data, This can lead to inflated performance scores driven by memorization rather than  rather than generalization. This risk is especially severe for static datasets, whose contents cannot evolve to stay ahead of future training sets.

To address this challenge, NotSoTiny is explicitly designed as a living benchmark. Tiny Tapeout releases between two and five new shuttles each year, and NotSoTiny leverages this continuous stream of fresh, real hardware designs to remain current. The benchmark will be periodically updated by integrating newly released shuttles while dropping older ones when necessary, ensuring that evaluation data stays ahead of what contemporary and future LLMs might have seen during training. Similar dynamic benchmarking strategies have proven effective in other domains, such as LiveBench for general LLM evaluation~\cite{white2025livebench}. While such dynamic benchmarking demands of maintenance and supervision, it is the only feasible solution in a field that evolves as fast as that of generative AI.

To quantitatively assess contamination, we adopt the Min-K~\cite{min-k} metric
which captures the average log-probability assigned by a model to the least likely K\% of tokens in a reference solution.
A low Min-K score means that the model found some parts of the solution quite unlikely, which is a strong hint that the solution was not seen during training. Conversely, a high Min-K indicates that even the most unexpected tokens were predicted with relatively high confidence, suggesting the model may have memorized the code. Min-K score can be thresholded to get a binary classifier for contamination. However, determining a threshold in practice is challenging. That’s why we follow the approach of Samuel et al.~\cite{samuel-etal-2025} to compare average Min-K values across different benchmarks, rather than thresholding them~\cite{min-k}.

The contamination study was performed on a diverse set of 11 LLMs released in a short period, July to December 2024, to ensure they were exposed to roughly the same data on the internet. The selection is diverse in terms of model size (7B to 405B parameters), family, and category (\textit{domain-specific}, \textit{coding-focused}, and \textit{general-purpose}). The models are Llama3.3-70B/ 3.1-405B~\cite{grattafiori2024llama}, Qwen2.5-32B/ 72B~\cite{yang2024qwen2}, Qwen2.5-Coder-7B/ 14B/ 32B~\cite{yang2024qwen2}, OpenCoder-8B~\cite{huang2024opencoder}, CodeV-QW~\cite{zhao2024codev}, Origen~\cite{origen}, HaVen-CodeQwen~\cite{haven} (using instruct variants if available). For all these models, we computed Min-K for NotSoTiny’s references and other benchmarks.

As noted by Wang et al.~\cite{VeriContaminated}, the Min-K may be sensitive to highly detailed prompts, as they significantly affect the likelihood of solutions. Therefore, to ensure a fair comparison, we compute only the Min-K of the reference module, without the context provided in the prompts. We excluded any large modules ($>$2000 tokens) so that NotSoTiny’s tasks are similar in size to those in RTLLM (mean of $\approx$400 tokens). 
We also exclude VerilogEval and VeriGen from the comparison because their modules are significantly smaller, approximately half the size and complexity of RTLLM (Table~\ref{tab:intro_stats}), and applying the same size-based filtering would discard most of NotSoTiny samples.

\begin{figure}[bt]
    \centering
    \includegraphics[width=0.99\linewidth]{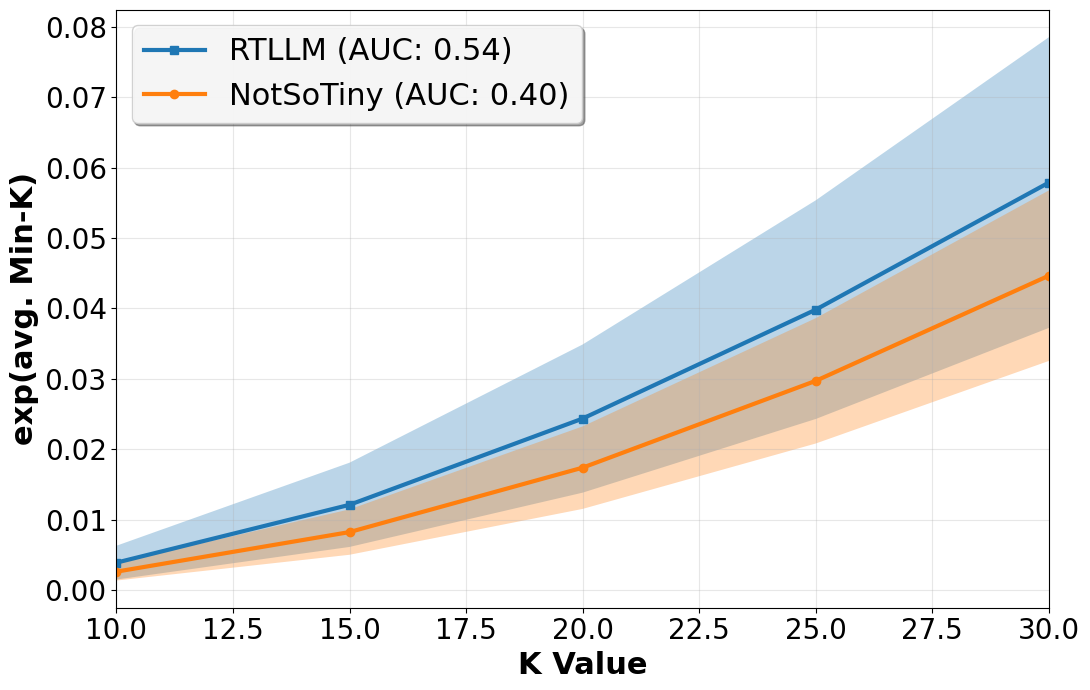}
    \caption{Average Min-K computed for different values of K (i.e., the K\% of tokens with lowest probability) around the value $K=20$ recommended by the authors of Min-K. Results are aggregated across all samples and models under study. Exponential is taken for easier interpretability with the Area Under the Curve (AUC), and thanks to its monotonicity, it doesn't affect the comparison. Higher values indicate higher chances of contamination.}
    \label{fig:benchmark_contamination}
    \vspace{-5pt}
\end{figure}

\begin{figure*}[t]
    \centering
    \includegraphics[width=0.95\textwidth]{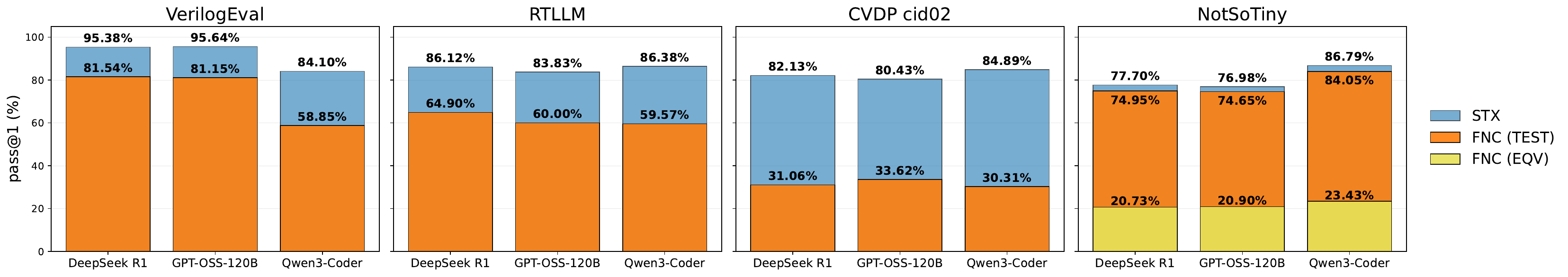}
     \caption{Syntax correctness (STX) and testbench correctness (FNC-Test) for three representative LLMs across four RTL benchmarks. Results show large drops from STX to FNC-Test on VerilogEval, RTLLM, and CVDP. Formal
equivalence correctness (FNC-EQV) is also included for the NotSoTiny benchmark
    }
    \label{fig:model_performance_comparision}
\end{figure*}

The final result can be seen in Figure~\ref{fig:benchmark_contamination}, which plots the average Min-K score as a function of K for NotSoTiny (aggregated over each NotSoTiny shuttle subset) versus the RTLLM benchmark. The difference is clear: RTLLM shows consistently higher Min-K values than NotSoTiny across the curve. In fact, the area under the Min-K curve for RTLLM is significantly greater, indicating a greater likelihood that models have seen those solutions during training (likely because RTLLM’s problems or their solutions were present in open-source code that made it into training data). This motivates the need for benchmarks supported by fully automated pipelines that permit periodic updates.


%% file: sections/4_Evaluations.tex
\section{Evaluations}
\label{sec:Evaluations}



To evaluate the functional correctness of LLM-generated RTL on NotSoTiny, we benchmark a broad set of models ranging from 7B to 1T parameters, including \textit{domain-specific}, \textit{coding-focused}, and \textit{general-purpose} models, at the end of this section. Models are chosen based on their previously reported performance at RTL tasks\footnote{\leaderboardurl}, and include Kimi K2~\cite{kimiteam2025kimik2openagentic}, DeepSeek R1 0528~\cite{deepseekai2025deepseekr1}, Qwen3 Coder 480B~\cite{yang2025qwen3technicalreport}, GPT-OSS 120B~\cite{gptoss120}, Qwen2.5 7B/14B/14B 1M/72B~\cite{yang2024qwen2}, Qwen2.5 Coder 32B~\cite{hui2024qwen25codertechnicalreport}, HaVen-CodeQwen~\cite{haven}, and OriGen~\cite{origen} (instruct versions used when available). 
These models were tested on NotSoTiny, and three of those also used to run prior RTL benchmarks (e.g., VerilogEval~\cite{pinckney2025revisiting}, RTLLM~\cite{lu2024rtllm}, CVDP~\cite{pinckney2025comprehensive}), for comparison purposes to evaluate the feasibility of using test-benches to check functionality in \S\ref{subsec:testbench}. Based on these results, formal verification is proposed and defined in \S\ref{subsec:formalverif}, with the full benchmarking results are presented and discuss in \S\ref{subsec:results}.

\subsection{Testbench Verification Analysis}\label{subsec:testbench}

\begin{figure}[t]
    \centering
    \includegraphics[width=1.0\linewidth]{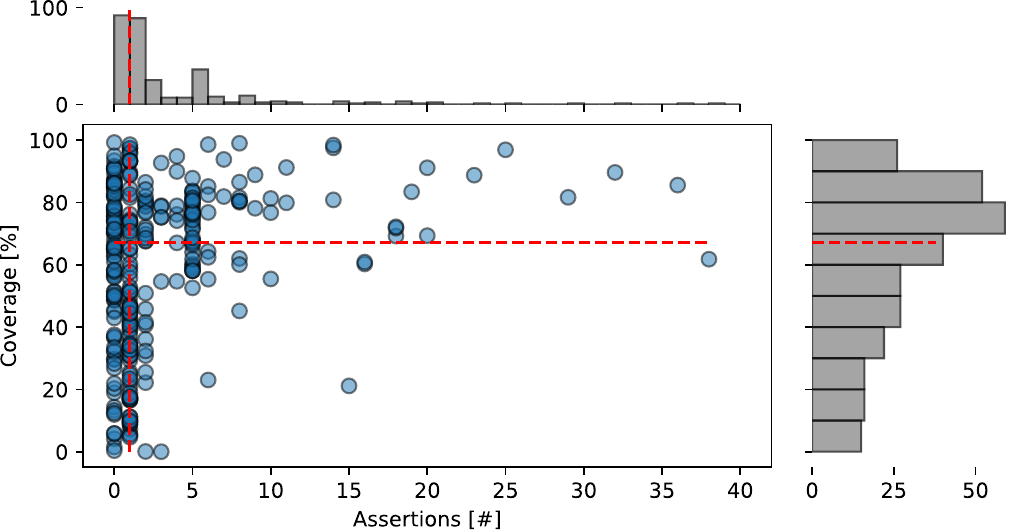}
    \caption{Coverage and assertion analysis of Tiny Tapeout testbenches. Each point represents a project, showing code coverage versus number of assertions. 
    }
    \label{fig:coverage}
    \vspace{-1.5em}
\end{figure}

For selecting a verification strategy for NotSoTiny, we first examine how syntax correctness (STX as Verilog compilation success) and simulation-based functional correctness (FNC-Test as testbench passing) behave across four benchmarks and three representative LLMs. Figure \ref{fig:model_performance_comparision} illustrates a consistently steep drop from STX to FNC-Test. This drop is particularly large in CVDP, where more than half of the syntactically correct solutions fail their testbenches. This indicates that functional errors are the main bottleneck in RTL generation, as observed in prior work~\cite{garciagasulla2025turtle}. However, NotSoTiny behaves differently, as the gap between STX and FNC-Test is negligible. Any generated code that is syntactically valid Verilog passes the associated testbenches 95–97\% of the time.

To further assess the relevance of Tiny Tapeout’s testbenches, we study two classical coverage dimensions: \textit{Reachability} and \textit{Observability}. 
\textit{Reachability} is measured through Synopsys VCS~\cite{Synopsys_VCS} code-coverage instrumentation, computing a consolidated score that aggregates \textit{line}, \textit{condition}, \textit{branch}, \textit{toggle}, and \textit{FSM} coverage metrics. 
This measures the extent to which the existing test stimulus can activate the design’s structural elements, a standard practice for identifying unexercised RTL regions.
\textit{Observability} is approximated by statically analyzing the Python \texttt{cocotb} tests and counting all occurrences of the \texttt{assert} keyword across the test files associated with each project, using assertion density as a coarse indicator of the testbench’s ability to detect and propagate design errors to an observable point, as emphasized in coverage-driven verification theory. 
Figure~\ref{fig:coverage} examines the quality of these native testbenches per project, in terms of code coverage and assertion density (ignoring approximately 2\% of the testbenches, featuring more than 30 assertions). 
The median coverage achieved by each project’s testbench is only about 67\%, and the median number of assertions is a mere 1 per project. In fact, only a small minority of the testbenches, approximately 26\%, contain more than one assertion and achieve at least 80\% code coverage. This means that, for most Tiny Tapeout projects, the test stimuli do not perform minimal checking of output correctness. These findings mirror well-known pitfalls in hardware verification: low reachability (not hitting all parts of the design) and low observability (not detecting errors at outputs) lead to test suites that miss bugs. 

\subsection{Formal Verification Analysis}\label{subsec:formalverif}

To overcome the limitations of simulation-based verification, we adopt \textit{formal verification} as our primary strategy for functional correctness. Specifically, we employ \textit{equivalence checking} to verify the equivalence of each generated module against the golden reference implementation. 

The verification flow is based on the Yosys~\cite{wolf2013yosys} suite, and it proceeds in three phases. In the first step, the golden RTL implementation and the LLM-generated candidate are both loaded into the tool and elaborated to ensure that the correct top-level module is selected and that both designs are represented in a consistent internal form. 

As a second step, the tool constructs a miter circuit: a composite structure that feeds identical inputs to both implementations, compares all corresponding outputs bit-wise through XNOR gates, and reduces these comparisons to a single equivalence signal. If the two designs are functionally identical, this signal remains high for all possible input combinations; any discrepancy drives it low, yielding a concrete counterexample. 

In the final step, we apply formal verification to the miter. We rely on two passes offered by Yosys: \texttt{equiv\_simple} and \texttt{equiv\_induct}. The \texttt{equiv\_simple} procedure performs combinational equivalence checking by formulating the question “Can any input assignment produce different outputs?” as a SAT instance over the miter circuit. If the SAT solver returns UNSAT, no such distinguishing input exists, and the designs are proven to be combinatorially equivalent across all possible inputs, providing exhaustive coverage unattainable with finite testbenches. To handle sequential behavior, we additionally use \texttt{equiv\_induct}, which extends verification to sequential logic using bounded $k$-induction. Sequential circuits maintain internal state across clock cycles, so single-step combinational checks are insufficient. The inductive method proceeds as follows: (1) \emph{Base case}, verify that both designs produce identical outputs from reset states; (2) \emph{Inductive step}, assume equivalence at cycle $t$, then prove equivalence at cycle $t+1$ for all possible inputs; (3) Repeat this reasoning for $k$ cycles (in our implementation, $k=30$). If the induction holds, the designs are proven equivalent for all reachable states within the bounded depth. This captures state-machine transitions, register initialization sequences, and multi-cycle behaviors that \texttt{equiv\_simple} alone cannot verify.

This formal approach provides a mathematical proof that the LLM-generated design and the golden reference produce identical outputs for all valid inputs and state sequences (up to the bounded depth), rather than checking only a finite sample of test vectors. It effectively eliminates false positives from incomplete testbenches while permitting architectural creativity, an LLM is free to implement, as long as the functional behavior is provably identical.

\subsubsection{Partition Coverage Metric}

While formal equivalence checking provides a binary pass/fail result, we introduce partition coverage as a fine-grained metric for partial credit. For the sake of equivalence checking, Yosys represents each design as a set of logic partitions. For the benchmark involving multiple partitions, we define \textit{Partition Coverage} as follows:

\begin{equation}
\text{Partition Coverage} = \frac{\sum_{n \in \text{equivalent partitions}} \text{partitions}(n)}{\sum_{m \in \text{all partitions}} \text{partitions}(m)} \times 100\%
\end{equation}

This metric quantifies \emph{how much} of the golden design's functionality the LLM successfully reproduced. For instance, if a task golden solution consists of three partitions and the LLM correctly implements two but fails on the third, traditional functionality checking methodologies would score this as a complete failure. Partition coverage, instead, reflects that a substantial portion of the design's logic was correctly implemented. 
This granularity enables distinguishing between generations that are ``completely wrong'' versus ``partially correct''. It also allows tracking which modules contribute to low coverage, and identifies systematic weaknesses.

\subsubsection{Advantages and Limitations of Equivalence Checking}

The proposed equivalence-checking-based functional evaluation pipeline offers a uniform, scriptable flow that applies to thousands of modules without manual test-vector authoring, while guaranteeing that no task-specific corner cases are missed, in the presence of a reference golden solution. Moreover, our experimentation relies on open-source tools from Yosys~\cite{wolf2013yosys}, ensuring reproducibility. However, formal equivalence checking also introduces some limitations. First, the inductive proof is bounded to $k=30$ cycles. Bugs that manifest only after more extended initialization sequences or deeper state-machine traversals may go undetected, and increasing $k$ incurs a high computational cost. Additionally, sequential equivalence checking assumes both designs start from comparable reset states. Different reset polarities or styles may be considered equivalent if they converge to the same functional behavior within a bounded depth.
Despite these limitations, our formal verification strategy provides a substantial improvement over prior RTL benchmarks that rely exclusively on syntax checking or limited simulation testbenches. The combination of exhaustive input-space exploration (via SAT solving), sequential reasoning (via bounded induction), ensures that functional bugs are detected with high confidence, while still permitting LLM-generated designs to explore diverse cycle-accurate microarchitectural solutions and allows us to check equivalence more efficiently than using induction alone.

\subsection{Model Performance}
\label{subsec:results}

\input{tables/main_table_results}

Table~\ref{tab:model_performance} summarizes the performance of 11 open-source LLMs, spanning different model families and parameter scales, evaluated on NotSoTiny using our formal equivalence–based metrics. Averaged across all Tiny Tapeout shuttles, these results reveal three key insights: 
First, model scale remains the strongest predictor of functional correctness. We observe clear stratification based on model size. Large-scale models (e.g., Qwen3-coder at 480B, DeepSeek R1 at 685B) reach EQV scores of 20--23\%, medium-scale models achieve 10--15\%, and small models (7B--13B) score below 5\%. Nonetheless, even the largest models fall far short of production-ready performance. Context length also plays an important role. Within the same parameter class, models with larger context windows consistently outperform those with shorter ones, reflecting the long-range contextual dependencies found in NotSoTiny.

Second, the gap between STX and EQV is substantial. 
While leading models exceed 85\% STX, EQV drops to roughly 20\%, a gap of more than 60 percentage points, illustrated in Figure~\ref{fig:model_performance_comparision}.
This highlights that \emph{functional correctness} for the generated RTL code remains elusive to modern LLMs. 

Finally, the partition coverage metric provides a more fine-grained evaluation of LLMs' ability to infer module behaviour from surrounding context. Even when a model fails full equivalence, it often reproduces 30–60\% of the design’s logic partitions correctly. This suggests that models may frequently solve simpler module partitions while struggling with heavy logic subsets. Partition coverage thus helps differentiate between generations that are entirely incorrect and those that partially capture the intended behavior.

%% file: tables/main_table_results.tex
\begin{table*}[h]
  \centering
   \caption{LLMs performance on NotSoTiny. The score with the final dataset is shown in the last column; the rest are the results obtained per shuttle. Includes parameter count, context length, syntax correctness (STX), formal equivalence correctness (EQV), and average partition coverage Cov$_\mu$. Best in \textbf{bold}. Using instruct versions of the models if available.}
  \label{tab:model_performance}
  \resizebox{\textwidth}{!}{
  \begin{tabular}{ l @{} rr *{8}{| c @{\hspace{2pt}} c @{\hspace{2pt}} c } }
  & & &  \multicolumn{3}{c}{\textbf{TT06}} & \multicolumn{3}{c}{\textbf{TT07}} &
  \multicolumn{3}{c}{\textbf{TT08}} & \multicolumn{3}{c}{\textbf{TT09}} & \multicolumn{3}{c}{\textbf{TT10 IHP-02}} & \multicolumn{3}{c}{\textbf{TT10 IHP-25a}} & \multicolumn{3}{c}{\textbf{TTSKY25a}} & \multicolumn{3}{c}{\textbf{NotSoTiny}}\\
  \cmidrule(lr){4-6} \cmidrule(lr){7-9} \cmidrule(lr){10-12} \cmidrule(lr){13-15} \cmidrule(lr){16-18}
  \cmidrule(lr){19-21} \cmidrule(lr){22-24} \cmidrule(lr){25-27}
  \textbf{Model} & $\uparrow$\textbf{Params} & \textbf{Context} & \textbf{STX} & \textbf{EQV} & \textbf{Cov$_\mu$} 
  & \textbf{STX} & \textbf{EQV} & \textbf{Cov$_\mu$}  & \textbf{STX} & \textbf{EQV} & \textbf{Cov$_\mu$} 
  & \textbf{STX} & \textbf{EQV} & \textbf{Cov$_\mu$}  & \textbf{STX} & \textbf{EQV} & \textbf{Cov$_\mu$} 
  & \textbf{STX} & \textbf{EQV} & \textbf{Cov$_\mu$}  & \textbf{STX} & \textbf{EQV} & \textbf{Cov$_\mu$} 
  & \textbf{STX} & \textbf{EQV} & \textbf{Cov$_\mu$} \\
  \midrule
Kimi-K2-0905 & 1000 B & 262,144 & 73.3 & 13.0 & 56.2 & 61.1 & 15.6 & 45.2 & 76.0 & 20.9 & 60.1 & 78.5 & 21.8 & 57.3 & 90.0 & 23.3 & 61.9 & 71.0 & 19.6 & 57.1 & 86.2 & 18.8 & 67.3 & 76.6 & 19.0 & 58.0 \\
DeepSeek-R1-0528 & 685 B & 163,840 & 76.9 & 14.6 & 59.8 & 61.1 & 20.6 & 46.2 & 75.6 & 19.7 & 59.5 & 83.1 & 23.8 & 59.7 & 85.3 & \textbf{24.7} & \textbf{60.0} & 72.4 & 20.7 & 59.5 & 89.5 & \textbf{21.1} & \textbf{69.8} & 77.7 & 20.7 & 59.2 \\
Qwen3-Coder-A35B & 480 B & 262,144 & \textbf{86.3} & \textbf{18.0} & \textbf{68.4} & \textbf{85.2} & \textbf{28.4} & \textbf{67.7} & \textbf{84.6} & \textbf{22.0} & \textbf{69.2} & \textbf{87.7} & 28.3 & \textbf{65.0} & \textbf{91.3} & 20.0 & 59.7 & \textbf{82.8} & \textbf{28.6} & \textbf{68.2} & \textbf{89.6} & 19.2 & 69.3 & \textbf{86.8} & \textbf{23.4} & \textbf{66.8} \\
gpt-oss-120b & 120 B & 131,072 & 80.9 & 7.4 & 63.0 & 59.9 & 19.2 & 45.0 & 73.3 & 19.8 & 56.2 & 88.1 & \textbf{31.8} & 64.9 & 86.0 & 23.3 & 60.3 & 67.3 & 18.2 & 52.8 & 83.4 & 16.6 & 63.4 & 77.0 & 20.9 & 57.9 \\
Qwen2.5-72B & 72 B & 32,768 & 64.4 & 11.7 & 44.8 & 50.6 & 12.8 & 36.1 & 47.8 & 5.6 & 32.0 & 76.1 & 22.4 & 51.1 & 68.0 & 22.7 & 41.7 & 58.3 & 15.1 & 44.1 & 72.0 & 12.7 & 51.6 & 62.4 & 14.7 & 43.0 \\
Qwen2.5-Coder-32B & 32 B & 32,768 & 61.1 & 7.8 & 41.8 & 47.5 & 10.2 & 33.0 & 46.8 & 5.9 & 30.9 & 74.2 & 17.5 & 48.2 & 62.0 & 13.3 & 36.0 & 55.8 & 15.4 & 40.9 & 73.9 & 12.8 & 54.3 & 60.2 & 11.8 & 40.7 \\
Qwen2.5-14B-1M & 14 B & 1,010,000 & 45.2 & 9.8 & 31.9 & 46.1 & 15.7 & 35.2 & 54.3 & 15.5 & 44.0 & 56.6 & 15.9 & 37.2 & 56.0 & 11.3 & 28.0 & 41.1 & 12.1 & 29.7 & 54.8 & 7.1 & 37.4 & 50.6 & 12.5 & 34.8 \\
Qwen2.5-14B & 14 B & 32,768 & 45.4 & 5.6 & 33.4 & 45.7 & 12.9 & 31.6 & 35.8 & 4.7 & 24.8 & 55.1 & 15.0 & 35.3 & 52.0 & 8.0 & 28.4 & 33.7 & 9.0 & 23.6 & 51.6 & 7.6 & 35.9 & 45.6 & 9.0 & 30.4 \\
HaVen-CodeQwen & 7 B & 65,536 & 32.0 & 2.8 & 14.9 & 27.1 & 3.4 & 14.4 & 25.3 & 1.7 & 14.3 & 48.6 & 7.2 & 18.7 & 40.0 & 10.7 & 22.1 & 22.2 & 0.4 & 11.2 & 38.3 & 3.2 & 20.2 & 33.4 & 4.2 & 16.6 \\
Qwen2.5-7B & 7 B & 32,768 & 13.3 & 0.7 & 8.3 & 17.4 & 6.2 & 12.6 & 12.9 & 2.9 & 8.2 & 14.8 & 3.8 & 10.3 & 12.0 & 7.3 & 9.0 & 8.9 & 1.7 & 6.1 & 16.5 & 1.7 & 10.3 & 13.7 & 3.5 & 9.3 \\
Origen-merged & 7 B & 4,096 & 7.6 & 4.1 & 6.9 & 5.3 & 2.7 & 4.6 & 3.1 & 1.0 & 2.6 & 10.0 & 5.8 & 8.7 & 12.0 & 7.3 & 10.6 & 4.2 & 2.2 & 3.3 & 1.3 & 0.0 & 0.9 & 6.2 & 3.3 & 5.4 \\
\bottomrule
\end{tabular}
}
\end{table*}

%% file: sections/5_Conclusions.tex
\section{Conclusion}
\label{sec:Conclusion}

This paper introduced NotSoTiny, a large, structurally rich, and contamination-resilient benchmark for evaluating LLMs on realistic RTL code generation. By framing tasks as contextual-module completion problems and validating outputs with scalable formal equivalence checking, NotSoTiny exposes the true difficulty of RTL generation for current LLMs. Our results show that even state-of-the-art models achieve only ~20\% functional correctness under formal verification. Our analyses further show that existing RTL benchmarks remain constrained by limited scale and are increasingly vulnerable to training-data contamination.

To enhance the visibility and accessibility, a
leaderboard\footnote{\leaderboardurl} is released and maintained under the TuRTLe framework~\cite{garciagasulla2025turtle}, as illustrated in Appendix~\ref{sec:appendix}.

%% file: sections/acknowledgment.tex
\section*{Acknowledgments}\label{sec:aknowledgment}
This work is supported by the AI4S fellowships awarded to Gokcen Kestor, Emanuele Parisi, Razine Moundir Ghorab, Cristian Gutierrez Gomez, and Miquel Albertí Binimelis as part of the “Generación D” initiative, Red.es, Ministerio para la Transformación Digital y de la Función Pública, for talent attraction (C005/24-ED CV1), funded by the European Union NextGenerationEU funds, through PRTR. Additionally, this contribution is also partially supported by the ELLIOT Grant funded by the European Union under grant agreement No. 101214398, and by the project PID2023-146511NB-I00 funded by the Spanish Ministry of Science, Innovation and Universities MCIU /AEI /10.13039/501100011033 and EU ERDF. We are grateful to the Operations department at BSC for their technical support.

Finally, we wish to extend our gratitude to Matt Venn and the entire Tiny Tapeout community, whose work and effort made our research possible. We also appreciate the time and effort contributed by Nikolas Belle, Dakota Barnes, Arnau Ayguadé, Bernat Homs, and Serik Perez in multiple discussions.

%% file: sections/appendix.tex
\newpage
\onecolumn
\section{Appendix: TuRTLe Leaderboard}\label{sec:appendix}

\begin{figure*}[h]
    \centering
    \includegraphics[width=\textwidth]{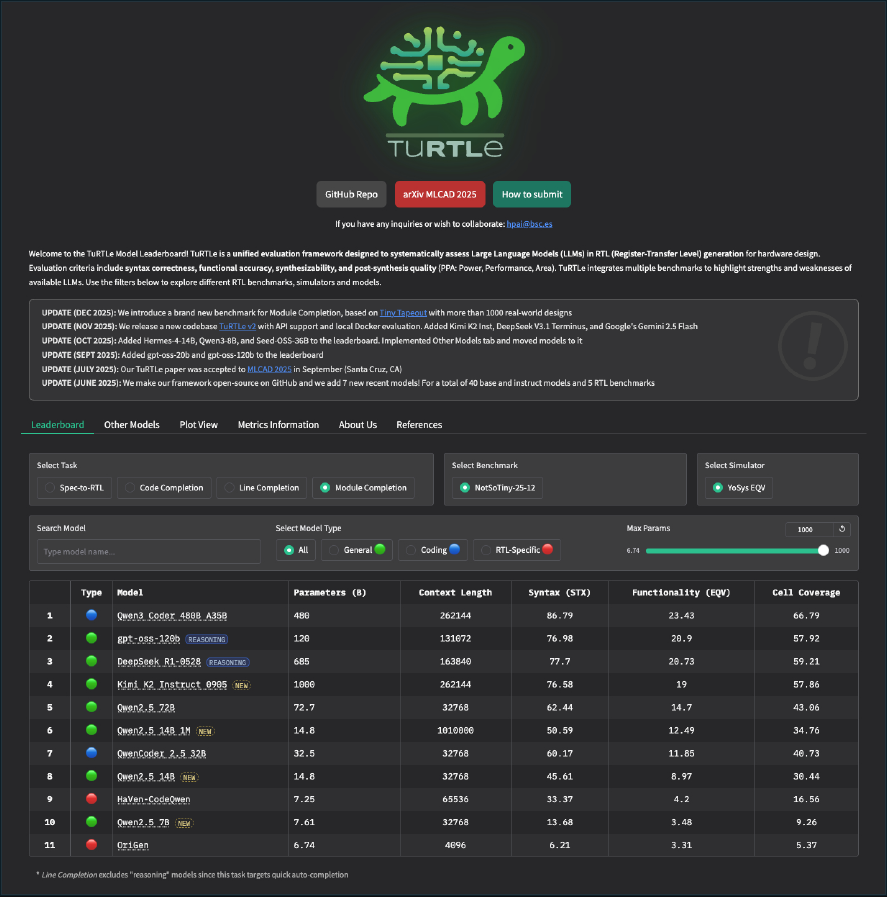}
     \caption{TuRTLe Leaderboard showing model performance on the NotSoTiny-25-12 benchmark for Module Completion. Available at: \url{https://huggingface.co/spaces/HPAI-BSC/TuRTLe-Leaderboard}.}
    \label{fig:leaderboard}
\end{figure*}

Figure~\ref{fig:leaderboard} presents the leaderboard for the NotSoTiny-25-12 benchmark, comprising 11 LLMs evaluated with the TuRTLe framework~\cite{garciagasulla2025turtle}. The selected models span general-purpose, code-specific, and RTL-specific, alongside some reasoning models. We use instruct versions when they are available.

%% file: references.bib
@inproceedings{
white2025livebench,
title={LiveBench: A Challenging, Contamination-Limited {LLM} Benchmark},
author={Colin White and Samuel Dooley and Manley Roberts and Arka Pal and Benjamin Feuer and Siddhartha Jain and Ravid Shwartz-Ziv and Neel Jain and Khalid Saifullah and Sreemanti Dey and Shubh-Agrawal and Sandeep Singh Sandha and Siddartha Venkat Naidu and Chinmay Hegde and Yann LeCun and Tom Goldstein and Willie Neiswanger and Micah Goldblum},
booktitle={The Thirteenth International Conference on Learning Representations},
year={2025},
url={https://openreview.net/forum?id=sKYHBTAxVa}
}

@misc{jin2025realbench,
	title        = {{RealBench: Benchmarking Verilog Generation Models with Real-World IP Designs}},
	author       = {Pengwei Jin and Di Huang and Chongxiao Li and Shuyao Cheng and Yang Zhao and Xinyao Zheng and Jiaguo Zhu and Shuyi Xing and Bohan Dou and Rui Zhang and Zidong Du and Qi Guo and Xing Hu},
	year         = 2025,
	url          = {https://arxiv.org/abs/2507.16200},
	eprint       = {2507.16200},
	archiveprefix = {arXiv},
	primaryclass = {cs.LG}
}

@inproceedings{garciagasulla2025turtle,
	title        = {{TuRTLe: A Unified Evaluation of LLMs for RTL Generation}},
	author       = {Garcia-Gasulla, Dario and Kestor, Gokcen and Parisi, Emanuele and Albertí-Binimelis, Miquel and Gutierrez, Cristian and Ghorab, Razine Moundir and Montenegro, Orlando and Homs, Bernat and Moreto, Miquel},
	year         = 2025,
	booktitle    = {2025 ACM/IEEE 7th Symposium on Machine Learning for CAD (MLCAD)},
	volume       = {},
	number       = {},
	pages        = {1--12},
	doi          = {10.1109/MLCAD65511.2025.11189228}
}

@inproceedings{dong2024datacontamination,
    title = "Generalization or Memorization: Data Contamination and Trustworthy Evaluation for Large Language Models",
    author = "Dong, Yihong  and
      Jiang, Xue  and
      Liu, Huanyu  and
      Jin, Zhi  and
      Gu, Bin  and
      Yang, Mengfei  and
      Li, Ge",
    editor = "Ku, Lun-Wei  and
      Martins, Andre  and
      Srikumar, Vivek",
    booktitle = "Findings of the Association for Computational Linguistics: ACL 2024",
    month = aug,
    year = "2024",
    address = "Bangkok, Thailand",
    publisher = "Association for Computational Linguistics",
    url = "https://aclanthology.org/2024.findings-acl.716/",
    doi = "10.18653/v1/2024.findings-acl.716",
    pages = "12039--12050"
}

@article{thakur2024verigen,
	title        = {{Verigen: A large language model for verilog code generation}},
	author       = {Thakur, Shailja and Ahmad, Baleegh and Pearce, Hammond and Tan, Benjamin and Dolan-Gavitt, Brendan and Karri, Ramesh and Garg, Siddharth},
	year         = 2024,
	journal      = {ACM Transactions on Design Automation of Electronic Systems},
	publisher    = {ACM New York, NY},
	volume       = 29,
	number       = 3,
	pages        = {1--31}
}

@inproceedings{liu2023verilogeval,
	title        = {{{VerilogEval:} Evaluating Large Language Models for Verilog Code Generation}},
	author       = {Liu, Mingjie and Pinckney, Nathaniel and Khailany, Brucek and Ren, Haoxing},
	year         = 2023,
	booktitle    = {2023 IEEE/ACM International Conference on Computer-Aided Design (ICCAD)},
	pages        = {1--8},
	organization = {IEEE}
}

@article{grattafiori2024llama,
	title        = {{The llama 3 herd of models}},
	author       = {Grattafiori, Aaron and Dubey, Abhimanyu and Jauhri, Abhinav and Pandey, Abhinav and Kadian, Abhishek and Al-Dahle, Ahmad and Letman, Aiesha and Mathur, Akhil and Schelten, Alan and Vaughan, Alex and others},
	year         = 2024,
	journal      = {arXiv preprint arXiv:2407.21783}
}

@article{pinckney2025revisiting,
	title        = {{Revisiting VerilogEval: A Year of Improvements in Large-Language Models for Hardware Code Generation}},
	author       = {Pinckney, Nathaniel and Batten, Christopher and Liu, Mingjie and Ren, Haoxing and Khailany, Brucek},
	year         = 2025,
	month        = {feb},
	journal      = {ACM Trans. Des. Autom. Electron. Syst.},
	publisher    = {Association for Computing Machinery},
	address      = {New York, NY, USA},
	doi          = {10.1145/3718088},
	issn         = {1084-4309},
	url          = {https://doi.org/10.1145/3718088},
	note         = {Just Accepted}
}

@inproceedings{lu2024rtllm,
	title        = {{RTLLM: An Open-Source Benchmark for Design RTL Generation with Large Language Model}},
	author       = {Lu, Yao and Liu, Shang and Zhang, Qijun and Xie, Zhiyao},
	year         = 2024,
	booktitle    = {Proceedings of the 29th Asia and South Pacific Design Automation Conference},
	location     = {Incheon, Republic of Korea},
	publisher    = {IEEE Press},
	series       = {ASPDAC '24},
	pages        = {722–727},
	doi          = {10.1109/ASP-DAC58780.2024.10473904},
	isbn         = 9798350393545,
	url          = {https://doi.org/10.1109/ASP-DAC58780.2024.10473904},
	numpages     = 6
}

@inproceedings{huang2024opencoder,
  title={Opencoder: The open cookbook for top-tier code large language models},
  author={Huang, Siming and Cheng, Tianhao and Liu, Jason Klein and Xu, Weidi and Hao, Jiaran and Song, Liuyihan and Xu, Yang and Yang, Jian and Liu, Jiaheng and Zhang, Chenchen and others},
  booktitle={Proceedings of the 63rd Annual Meeting of the Association for Computational Linguistics (Volume 1: Long Papers)},
  pages={33167--33193},
  year={2025}
}

@misc{yang2024qwen2,
	title        = {{Qwen2.5 Technical Report}},
	author       = {Yang, An and Yang, Baosong and Zhang, Beichen and Hui, Binyuan and Zheng, Bo and Yu, Bowen and Li, Chengyuan and Liu, Dayiheng and Huang, Fei and Wei, Haoran and others},
	year         = 2025,
	url          = {https://arxiv.org/abs/2412.15115},
	eprint       = {2412.15115},
	archiveprefix = {arXiv},
	primaryclass = {cs.CL}
}

@inproceedings{haven,
	title        = {{HaVen: Hallucination-Mitigated LLM for Verilog Code Generation Aligned with HDL Engineers}},
	author       = {Yang, Yiyao and Teng, Fu and Liu, Pengju and Qi, Mengnan and Lv, Chenyang and Li, Ji and Zhang, Xuhong and He, Zhezhi},
	year         = 2025,
	booktitle    = {2025 Design, Automation \& Test in Europe Conference (DATE)},
	volume       = {},
	number       = {},
	pages        = {1--7},
	doi          = {10.23919/DATE64628.2025.10993072}
}

@misc{zhao2024codev,
	title        = {{CodeV: Empowering LLMs with HDL Generation through Multi-Level Summarization}},
	author       = {Yang Zhao and Di Huang and Chongxiao Li and Pengwei Jin and Muxin Song and Yinan Xu and Ziyuan Nan and Mingju Gao and Tianyun Ma and Lei Qi and Yansong Pan and Zhenxing Zhang and Rui Zhang and Xishan Zhang and Zidong Du and Qi Guo and Xing Hu},
	year         = 2025,
	url          = {https://arxiv.org/abs/2407.10424},
	eprint       = {2407.10424},
	archiveprefix = {arXiv},
	primaryclass = {cs.PL}
}

@misc{deepseekai2025deepseekr1,
	title        = {{DeepSeek-R1: Incentivizing Reasoning Capability in LLMs via Reinforcement Learning}},
	author       = {DeepSeek-AI},
	year         = 2025,
	url          = {https://arxiv.org/abs/2501.12948},
	eprint       = {2501.12948},
	archiveprefix = {arXiv},
	primaryclass = {cs.CL}
}

@misc{pinckney2025comprehensive,
	title        = {{Comprehensive Verilog Design Problems: A Next-Generation Benchmark Dataset for Evaluating Large Language Models and Agents on RTL Design and Verification}},
	author       = {Nathaniel Pinckney and Chenhui Deng and Chia-Tung Ho and Yun-Da Tsai and Mingjie Liu and Wenfei Zhou and Brucek Khailany and Haoxing Ren},
	year         = 2025,
	journal      = {arXiv preprint arXiv:2506.14074},
	url          = {https://arxiv.org/abs/2506.14074},
	eprint       = {2506.14074},
	archiveprefix = {arXiv},
	primaryclass = {cs.LG}
}

@inproceedings{samuel-etal-2025,
	title        = {{Towards Data Contamination Detection for Modern Large Language Models: Limitations, Inconsistencies, and Oracle Challenges}},
	author       = {Samuel, Vinay  and Zhou, Yue  and Zou, Henry Peng},
	year         = 2025,
	month        = jan,
	booktitle    = {Proceedings of the 31st International Conference on Computational Linguistics},
	publisher    = {Association for Computational Linguistics},
	address      = {Abu Dhabi, UAE},
	pages        = {5058--5070},
	url          = {https://aclanthology.org/2025.coling-main.338/},
	editor       = {Rambow, Owen  and Wanner, Leo  and Apidianaki, Marianna  and Al-Khalifa, Hend  and Eugenio, Barbara Di  and Schockaert, Steven}
}

@software{Snyder_Verilator,
	title        = {{Verilator}},
	author       = {Snyder, Wilson and Wasson, Paul and Galbi, Duane and {et al}},
	year         = 2025,
	url          = {https://github.com/verilator/verilator},
	license      = {["LGPL-3.0-only", "Artistic-2.0"]}
}

@software{Synopsys_VCS,
	title        = {{VCS: Functional Verification Solution}},
	author       = {Synopsys, Inc.},
	url          = {https://www.synopsys.com/verification/simulation/vcs.html},
	license      = {Commercial}
}

@article{pan2025survey,
	title        = {{A Survey of Research in Large Language Models for Electronic Design Automation}},
	author       = {Pan, Jingyu and Zhou, Guanglei and Chang, Chen-Chia and Jacobson, Isaac and Hu, Jiang and Chen, Yiran},
	year         = 2025,
	month        = feb,
	journal      = {ACM Trans. Des. Autom. Electron. Syst.},
	publisher    = {Association for Computing Machinery},
	address      = {New York, NY, USA},
	volume       = 30,
	number       = 3,
	doi          = {10.1145/3715324},
	issn         = {1084-4309},
	url          = {https://doi.org/10.1145/3715324},
	issue_date   = {May 2025},
	articleno    = 34,
	numpages     = 21
}

@article{he2025large,
	title        = {{Large Language Models for EDA: Future or Mirage?}},
	author       = {He, Zhuolun and Pu, Yuan and Wu, Haoyuan and Qiu, Tairu and Yu, Bei},
	year         = 2025,
	month        = oct,
	journal      = {ACM Trans. Des. Autom. Electron. Syst.},
	publisher    = {Association for Computing Machinery},
	address      = {New York, NY, USA},
	volume       = 30,
	number       = 6,
	doi          = {10.1145/3736167},
	issn         = {1084-4309},
	url          = {https://doi.org/10.1145/3736167},
	issue_date   = {November 2025},
	articleno    = 90,
	numpages     = 53
}

@article{tinytapeout,
	title        = {{Tiny Tapeout: A shared silicon tape out platform accessible to everyone}},
	author       = {Venn, Matt},
	year         = 2024,
	journal      = {IEEE Solid-State Circuits Magazine},
	volume       = 16,
	number       = 2,
	pages        = {20--29},
	doi          = {10.1109/MSSC.2024.3381097}
}

@inproceedings{VeriContaminated,
	title        = {{VeriContaminated: Assessing LLM-Driven Verilog Coding for Data Contamination}},
	author       = {Wang, Zeng and Shao, Minghao and Bhandari, Jitendra and Mankali, Likhitha and Karri, Ramesh and Sinanoglu, Ozgur and Shafique, Muhammad and Knechtel, Johann},
	year         = 2025,
	booktitle    = {2025 IEEE International Conference on LLM-Aided Design (ICLAD)},
	volume       = {},
	number       = {},
	pages        = {117--123},
	doi          = {10.1109/ICLAD65226.2025.00017}
}

@inproceedings{min-k,
	title        = {{DETECTING PRETRAINING DATA FROM LARGE LANGUAGE MODELS}},
	author       = {Shi, Weijia and Ajith, Anirudh and Xia, Mengzhou and Huang, Yangsibo and Liu, Daogao and Blevins, Terra and Chen, Danqi and Zettlemoyer, Luke},
	year         = 2024,
	booktitle    = {12th International Conference on Learning Representations, ICLR 2024}
}

@inproceedings{origen,
	title        = {{OriGen: Enhancing RTL Code Generation with Code-to-Code Augmentation and Self-Reflection}},
	author       = {Cui, Fan and Yin, Chenyang and Zhou, Kexing and Xiao, Youwei and Sun, Guangyu and Xu, Qiang and Guo, Qipeng and Liang, Yun and Zhang, Xingcheng and Song, Demin and Lin, Dahua},
	year         = 2025,
	booktitle    = {Proceedings of the 43rd IEEE/ACM International Conference on Computer-Aided Design},
	location     = {Newark Liberty International Airport Marriott, New York, NY, USA},
	publisher    = {Association for Computing Machinery},
	address      = {New York, NY, USA},
	series       = {ICCAD '24},
	doi          = {10.1145/3676536.3676830},
	isbn         = 9798400710773,
	url          = {https://doi.org/10.1145/3676536.3676830},
	articleno    = 99,
	numpages     = 9
}

@misc{yang2025qwen3technicalreport,
	title        = {{Qwen3 Technical Report}},
	author       = {An Yang and Anfeng Li and Baosong Yang and {et al}},
	year         = 2025,
	url          = {https://arxiv.org/abs/2505.09388},
	eprint       = {2505.09388},
	archiveprefix = {arXiv},
	primaryclass = {cs.CL}
}

@misc{gptoss120,
	title        = {{gpt-oss-120b \& gpt-oss-20b Model Card}},
	author       = {OpenAI team},
	year         = 2025,
	url          = {https://arxiv.org/abs/2508.10925},
	eprint       = {2508.10925},
	archiveprefix = {arXiv},
	primaryclass = {cs.CL}
}

@misc{kimiteam2025kimik2openagentic,
	title        = {{Kimi K2: Open Agentic Intelligence}},
	author       = {Kimi Team},
	year         = 2025,
	url          = {https://arxiv.org/abs/2507.20534},
	eprint       = {2507.20534},
	archiveprefix = {arXiv},
	primaryclass = {cs.LG}
}

@software{eric_zhu_2017_290602,
  author    = {Eric Zhu and
               Vadim Markovtsev},
  doi       = {10.5281/zenodo.290602},
  month     = feb,
  publisher_e = {Zenodo},
  title     = {ekzhu/datasketch: First stable release},
  url       = {https://doi.org/10.5281/zenodo.290602},
  version   = {v1.0.0},
  year      = 2017
}

@book{Leskovec_Rajaraman_Ullman_2014, place={Cambridge}, edition={2}, title={Mining of Massive Datasets}, publisher={Cambridge University Press}, author={Leskovec, Jure and Rajaraman, Anand and Ullman, Jeffrey David}, year={2014}}

@misc{bigcode-dedup,
  author = {Chenghao Mou},
  year = {2023},
  title = {Large-scale Near-deduplication Behind BigCode},
  url = {https://huggingface.co/blog/dedup},
}

@misc{chen2021evaluatinglargelanguagemodels,
      title={Evaluating Large Language Models Trained on Code}, 
      author={Mark Chen and Jerry Tworek and Heewoo Jun and Qiming Yuan and Henrique Ponde de Oliveira Pinto and Jared Kaplan and Harri Edwards and Yuri Burda and Nicholas Joseph and Greg Brockman and Alex Ray and Raul Puri and Gretchen Krueger and Michael Petrov and Heidy Khlaaf and Girish Sastry and Pamela Mishkin and Brooke Chan and Scott Gray and Nick Ryder and Mikhail Pavlov and Alethea Power and Lukasz Kaiser and Mohammad Bavarian and Clemens Winter and Philippe Tillet and Felipe Petroski Such and Dave Cummings and Matthias Plappert and Fotios Chantzis and Elizabeth Barnes and Ariel Herbert-Voss and William Hebgen Guss and Alex Nichol and Alex Paino and Nikolas Tezak and Jie Tang and Igor Babuschkin and Suchir Balaji and Shantanu Jain and William Saunders and Christopher Hesse and Andrew N. Carr and Jan Leike and Josh Achiam and Vedant Misra and Evan Morikawa and Alec Radford and Matthew Knight and Miles Brundage and Mira Murati and Katie Mayer and Peter Welinder and Bob McGrew and Dario Amodei and Sam McCandlish and Ilya Sutskever and Wojciech Zaremba},
      year={2021},
      eprint={2107.03374},
      archivePrefix={arXiv},
      primaryClass={cs.LG},
      url={https://arxiv.org/abs/2107.03374}, 
}

@inproceedings{wolf2013yosys,
  title        = {{Yosys-a free Verilog synthesis suite}},
  author       = {Wolf, Clifford and Glaser, Johann and Kepler, Johannes},
  year         = 2013,
  booktitle    = {Proceedings of the 21st Austrian Workshop on Microelectronics (Austrochip)},
  volume       = 97
}

@misc{hui2024qwen25codertechnicalreport,
      title={Qwen2.5-Coder Technical Report}, 
      author={Binyuan Hui and Jian Yang and Zeyu Cui and Jiaxi Yang and Dayiheng Liu and Lei Zhang and Tianyu Liu and Jiajun Zhang and Bowen Yu and Keming Lu and Kai Dang and Yang Fan and Yichang Zhang and An Yang and Rui Men and Fei Huang and Bo Zheng and Yibo Miao and Shanghaoran Quan and Yunlong Feng and Xingzhang Ren and Xuancheng Ren and Jingren Zhou and Junyang Lin},
      year={2024},
      eprint={2409.12186},
      archivePrefix={arXiv},
      primaryClass={cs.CL},
      url={https://arxiv.org/abs/2409.12186}, 
}
